\documentclass[a4paper,12pt]{article}
\usepackage{jcappub}
\usepackage{amsthm,graphicx,multirow}
\usepackage{epsfig}
\usepackage{latexsym, amssymb} 
\usepackage{amsmath}
\usepackage{comment}
\usepackage{hyperref}
\usepackage[utf8]{inputenc}

{
{

\begin{document}

\title{The `unitarity problem' of Higgs inflation in the light of collapse dynamics}%

\author[a]{Suratna Das}%

\affiliation[a]{Department of Physics, Indian Institute of Technology, Kanpur 208016, India.}

\emailAdd{suratna@iitk.ac.in}

\abstract{Higgs inflation scenario is one of the most compelling models of inflation at present time. It not only explains the observed data well, but also provides means to include the inflaton field within the well understood Standard Model of particle physics, without invoking any need for its extension. Despite this, due to the requirement of large non-minimal coupling to the curvature scalar of the inflaton field, or in this case the Higgs field, this model suffers from a problem often called as the `unitarity' or the `naturalness' problem. On the other hand, to address the longstanding `interpretational issue' of quantum to classical transition of the primordial modes, the collapse dynamics of quantum mechanics has recently been included into the inflationary mechanism. We show that inclusion of such collapse mechanism in Higgs inflation helps alleviate the `unitarity problem' to a great deal. 
 }

\maketitle
\section{Introduction}

Large field inflationary models with simple quadratic and quartic self potentials, first proposed by Linde in his {\it chaotic inflation} scenario in 1983 \cite{Linde:1983gd}, are increasingly at odds with the observations, and this fact raises concerns for building concrete particle physics models of inflationary dynamics. The obstacle with these potentials is twofold. First of all, it was known for a long time that the usual assumption of minimally coupled inflaton field calls for severe fine-tuning of the self-couplings of the inflaton field in order to explain the observed density fluctuations, e.g. for quartic inflaton potential, $\lambda \phi^4$, the self-coupling constant is to be tuned to $\lambda\sim\mathcal{O}\left(10^{-13}\right)$ \cite{Linde:1983gd}. It was first observed by Spokoiny in 1984 \cite{Spokoiny:1984bd} and was investigated in full detail by Fakir and Unruh in 1990 \cite{Fakir:1990eg}, that a non-minimal curvature coupling of the inflaton field, such as
\begin{eqnarray}
\mathcal{L}=\left(\frac{1}{16\pi G}+\frac12\xi \phi^2\right)R+\frac12\nabla_\mu\phi\nabla^\mu\phi-V(\phi),
\end{eqnarray}
would help ease the severe fine-tuning of the inflaton quartic self-coupling and would yield a sufficiently long inflationary period which was in agreement with the then available observational data. It was later realised that such monomial potentials yield way too large tensor-to-scalar ratios to be in accordance with the present observations. The single field monomial potentials in the large field inflationary models,
\begin{eqnarray}
V(\phi)=\lambda M_{\rm Pl}^4\left(\frac{\phi}{M_{\rm Pl}}\right)^n,
\end{eqnarray}
give rise to the scalar spectral index as $n_s-1\approx -2(n+2)/(4N_*+n)$ and the tensor-to-scalar ratio as $r\approx 16n/(4N_*+n)$, where $N_*$ is the number of efolds when the observed pivot scale had crossed the horizon-scale during inflation \cite{Ade:2015lrj}. Considering $N_*\approx 60$ this yields $r\sim 0.26$ for $n=4$ and $r\sim 0.13$ for $n=2$, where the present observational upper bound on tensor-to-scalar ratio is $r\lesssim0.07$ \cite{Array:2015xqh} which disfavours simple quartic and quadratic potentials. It was first observed by Bezrukov and Shaposhnikov in the year 2008 \cite{Bezrukov:2007ep} that due to such non-minimal curvature coupling the scalar spectral index and the tensor-to-scalar ratio in the case of quartic coupling would become $n_s-1\simeq-8(4N_*+9)/(4N+3)^2$ and $r\approx 192/(4N_*+3)^2$ respectively, and hence for $N_*=60$, such a non-minimal model would yield $r\simeq0.0033$, which is well within the observational upper bound.\footnote{Recently, it has been pointed out that a 4$^{\rm th}$ order polynomial potential, a combination of quadratic, cubic and quartic potentials, with canonical minimal coupling to gravity can still be in accordance with present observations \cite{Musoke:2017frr}.} 

It is to note that the value of the non-minimal coupling constant $\xi$ plays no direct role either in bringing down the tensor-to-scalar ratio by two orders of magnitude or in determining the scalar spectral index, whereas it plays a major role in attenuating the severe fine-tuning of the self-coupling constant to match the observed amplitude of the scalar spectrum. As we will see later, the scalar power amplitude in the minimal setup is directly proportional to the self-coupling constant $\lambda$, whereas introducing the non-minimal curvature coupling the scalar power spectrum, calculated in the Einstein frame, turns out to be proportional to $\lambda/\xi^2$. Hence it is obvious that a large value of $\xi$ would help alleviating the fine-tuning constraint of the self-coupling constant. It is noted in \cite{Bezrukov:2007ep} that one requires $\xi$ as high as $\mathcal{O}(10^4)$ to get $\lambda\sim\mathcal{O}(10^{-1})$. Moreover, it was also noted that considering this inflaton field as the Standard Model (SM) Higgs field, the electroweak symmetry breaking scale $v$ does not affect the inflationary dynamics, and hence the Higgs field, the only known scalar in nature so far, itself can play the role of inflaton \cite{Bezrukov:2007ep}. This yields a minimal particle physics setup where the inflaton field can be successfully integrated into the SM without invoking any further extension of it. This Higgs inflation model gains popularity after the release of the Planck observation data, as it places this model in the `sweet spot' of the $n_s-r$ plot yielding it to be one of the most compelling models of inflation \cite{Ade:2015lrj}. 

The non-observability of the non-minimal coupling $\xi$ in the inflationary observables provides one the freedom to choose any value of it as per one's convenience, even as large as of the order of $10^4$. But, it was noted in \cite{Burgess:2009ea, Barbon:2009ya, Burgess:2010zq} that the non-renormalizability of the added non-minimal curvature coupling term to the theory renders a cut off scale at $\Lambda=M_{\rm Pl}/\xi$, beyond which the theory ceases to be remained valid. With $\xi$ as large as $10^4$ this cut off scale is much lower than the GUT scale, which is presumably  the scale of inflation according to the current observational upper bound on $r$ \cite{Ade:2015lrj}. Above all, the value of $\Lambda$ is much smaller than the value of the Higgs field during inflation. Hence it is legitimate to consider higher order terms of the Higgs field in the potential (like $(H^\dagger H)^n/\Lambda^{2n-4}$, $n\geq 3$ in the small field limit), which in turn would potentially destroy the slow-roll regime.\footnote{The effect of such higher order correction terms to the Higgs potential during inflation is still being investigated in the present literature \cite{Bezrukov:2017dyv, Fumagalli:2016lls}.} This problem is often termed as the `unitarity problem' or the `naturalness problem' of the Higgs inflation scenario. This problem was counteracted in \cite{Bezrukov:2010jz} (see \cite{Barvinsky:2009ii} for a renormalization group improved analysis), by considering a toy model of a single scalar field non-minimally coupled to the curvature scalar, where it was shown that the cut off scale depends on the background value of the scalar field and during inflation, in particular, it coincides with the Planck scale in the Einstein frame which is much higher than the Hubble rate during that time.\footnote{There are other analysis, like \cite{Calmet:2013hia, Escriva:2016cwl, Prokopec:2014iya}, which determined the cut-off scale as the onset of strong coupling regime in an attempt to evade the `unitarity' problem of Higgs inflation.} Furthermore, an assumption of asymptotic shift symmetry allows one to preserve the flatness of the potential under radiative corrections during inflation. But it was pointed out in \cite{Giudice:2010ka} that the issue with the large $\xi$ raises further concerns because including the gauge fields in this simple toy model examined in \cite{Bezrukov:2010jz} brings down the cut off scale during inflation at $M_{\rm Pl}/\sqrt{\xi}$ which is parametrically close to the scales involved in inflation. Hence to achieve a lower value of $\xi$, another scalar $\sigma$ of mass of the cut off scale $M_{\rm Pl}/\sqrt{\xi}$ was introduced into the theory in \cite{Giudice:2010ka}. However, such a framework introduces more unknown parameters to the theory (unknown couplings of the newly introduced scalar field $\sigma$), now to alleviate the strong coupling constraint on the non-minimal coupling, and also goes beyond the philosophy of the `minimalistic approach' of Higgs inflation. Thus, within the framework of Higgs inflation scenario, the required large value of $\xi$ remains to be of concern till date. It is understandable that one requires to bring in new parameters in the theory in order to tame the large values of $\xi$, but then it would be preferred if the new introduced parameters help serve the purpose of resolving other long-standing issues of inflationary paradigm apart from addressing the unitarity problem of Higgs inflation alone. 

On the other hand, the classical nature of the primordial perturbations, which are considered to be generated quantum mechanically during inflation, raises an issue of quantum to classical transition of these primordial modes, which is still being analysed in the literature. This problem can be cast as a more serious form of the `Measurement problem'  of quantum mechanics. Hence, the standard frameworks which are often being employed in quantum mechanics to deal with such a problem, such as mechanisms of decoherence \cite{Kiefer:2008ku, Kiefer:2006je, Kiefer:1998qe, Polarski:1995jg} and Bohmian mechanics \cite{Goldstein:2015mha}, have both been considered in good detail in order to address this long standing `interpretational issue' associated with the inflationary mechanism. Besides, quantum mechanical models of spontaneous wave-function collapse, which are observationally falsifiable in the regime which is dubbed the mesoscopic scale (a domain which is larger than the microscopic scales but smaller than macroscopic scales, and is still unexplored by observations), are being devised over the past three decades to address the `Measurement problem'  of quantum mechanics \cite{Bassi:2012bg}.  Quantum mechanical collapse dynamics modify the generic Schr\"{o}dinger equation by adding non-linear and stochastic terms to manifest the breakdown of the superposition of wavefunctions and the probabilistic outcome of the measurement. The advantage of collapse dynamics over decoherence and Bohmian mechanics is that the modified quantum dynamics leave distinguishable imprints on the observables. The Continuous Spontaneous Localization (CSL) model, first proposed by Pearle in 1989 \cite{Pearle:1988uh} and then by Ghirardi, Pearle and Rimini in 1990 \cite{Ghirardi:1989cn}, is favoured over its preceding models, like GRW (Ghirardi, Rimini, Weber model \cite{Ghirardi:1985mt}) and QMUPL (Quantum Mechanics with Universal Position Localization model \cite{Diosi:1988uy}), as it overcomes some of the previous difficulties, like applying a collapse model to a system of identical particles and accommodating a random field as a function of both space and time in order to collapse the wavefunction in space. Thus the CSL model remains to be the most advanced collapse model to date. Recently, the CSL model, has also been employed to inflationary perturbation theory in order to address the `macro-objectification' of the primordial modes \cite{Canate:2012ua, Landau:2011aa, Martin:2012pea, Das:2013qwa, Das:2014ada, Banerjee:2016klg}, which leaves distinguishable features in the cosmological observations \footnote{Recently it has been pointed out in \cite{Martin:2018zbe, Martin:2018lin}  that in case of decoherence the interaction of the primordial quantum modes with its environment leaves distinguishable imprints on primordial power spectrum and trispectrum.}.

It is to note that the CSL model has been implemented into inflationary dynamics in two different ways. According to the analysis done by Sudarsky and collaborators \cite{Canate:2012ua, Landau:2011aa} the CSL collapse dynamics helps seed the primordial inhomogeneities in a homogenous cosmological background which also explains the quantum to classical transition of these modes in later stages (we will call this approach of implementing CSL dynamics into inflationary mechanism as `CSL mechanism I'). In a different approach, which was first proposed by Martin and his collaborators in \cite{Martin:2012pea} and later developed in \cite{Das:2013qwa, Das:2014ada}, CSL mechanism has been integrated into the inflationary dynamics only to explain the quantum to classical transition of the primordial perturbations which were treated as the quantum fluctuations of the inflaton field, as it is done in the standard paradigm of inflation (we will call this approach of implementing CSL dynamics into inflationary mechanism as `CSL mechanism II'). Both these approaches lead to different predictions of the inflationary observables. One such striking difference is that while the second approach yields the tensor-to-scalar ratio just the same as in the standard picture, the first approach does not produce any tensor modes, thus its key signature would be the absence of the tensor modes in CMB observations \cite{Leon:2011ca}. Recently, the Higgs inflationary scenario has been analysed within the framework of this first approach by Sudarsky and collaborators and it has been shown that the non-minimal coupling $\xi$ can be brought down to less than one \cite{Rodriguez:2017rjh}. The purpose of this article would be to analyse both these approaches within the framework of Higgs inflation and then to qualitatively compare their outcomes. 

The rest of the article is organised as follows. In Section~(\ref{models}) we will briefly discuss the two approaches to implement the CSL dynamics into inflationary mechanism. In Section~(\ref{observables}) we will compare how introduction of collapse dynamics would help alleviate the `unitarity problem' of Higgs inflation scenario. Then in Section~(\ref{conclusion}) we will compare the results obtained in the previous section and then will conclude. 
 
\section{Brief overviews of both the approaches to implement CSL dynamics into inflationary paradigm}
\label{models}

In this section we provide very brief overviews of both the approaches to implement the CSL dynamics into the inflationary paradigm in order to mark the differences between these two approaches which lead to different observational predictions. 
\subsection{CSL Mechanism I}

To provide a brief description of the approach advocated by Sudarsky and collaborators we would follow \cite{Canate:2012ua, Rodriguez:2017rjh}. In this approach, the background spacetime structure is to be treated classically where the quantum perturbations of the inflaton field evolves in an adiabatic vacuum state. Collapse of these inflaton quantum perturbations evolving under the CSL dynamics leads to a change in the quantum state which then yields the primordial inhomogeneities in the background metric. 

The evolution of a state vector under a modified Schr\"{o}dinger equation that has been considered in \cite{Canate:2012ua} is 
\begin{eqnarray}
\left|\phi,t\rangle_w\right.={\rm T}\exp\left(-\int_0^tdt'\left\{i\hat{H}+\frac{1}{4\lambda_c}[w(t')-2\lambda_c\hat{A}]^2\right\}\right)\left|\phi,0\rangle\right.,
\end{eqnarray}
where $\rm{T}$ is the `time ordering' operator, $\hat{A}$ is the collapse operator which determines the preferred basis into which the underlying collapse dynamics would drive an initial arbitrary state, $\lambda_c$ is the collapse parameter, and $w(t)$ is the white noise random function which is responsible for the stochastic evolution of the system. The dimension of the collapse parameter $\lambda_c$ would depend upon how the collapse operator $\hat{A}$ is being chosen, as we will state below. 

One important feature which is common to both the approaches and which will play the most important role in applying the collapse dynamics in Higgs inflation scenario is that {\it all the observables are to be calculated at the end of inflation}. In a generic inflationary scenario the perturbation modes are calculated at the horizon-crossing during inflation, because these modes `freeze' on superhorizon scales and they evolve only with the background scale factor thereafter. However, the reason provided in this approach for calculating the imprints of inflaton fluctuations at the end of inflation is due to the assumption that the fluctuating fields provides seeds of structure in the form of  inhomogeneities and anisotropies in the ordinary matter distribution by the reheating conversion process at the end of inflation. To serve this purpose one requires to calculate the conformal time at the beginning and end of inflation, denoted by $\mathcal{T}$ and $\tau$ respectively.

The observables in this scenario is then directly calculated in terms of the temperature fluctuations in the CMB sky, which can be expanded in spherical harmonics in the last scattering surface as 
\begin{eqnarray}
\frac{\Delta T}{\overline{T}} =\sum_{\ell, m}\alpha_{\ell m}Y_{\ell m}, 
\end{eqnarray}
where the coefficients $\alpha_{\ell m}$ can be calculated  as it appears in the two-point correlation function or the power spectrum of these temperature fluctuations as 
\begin{eqnarray}
\overline{|\alpha_{\ell m}|^2}=(4\pi c)^2\int_0^\infty \frac{dk}{k}j_\ell^2(kR_D)\frac{\overline{\langle\hat{\pi}(\mathbf{k}, \eta) \rangle^2}}{k}, \quad\quad \left(c\equiv -\frac{4\pi G}{3a}\phi'_0(\eta) \right),
\end{eqnarray}
where the collapse operator is chosen as the field conjugate momentum $\hat{A}=\hat{\pi}=a\delta\hat{\phi}'$. If $\overline{\langle\pi\rangle^2}=\alpha k$ with a constant $\alpha$, then this mechanism yields a scale invariant spectrum,
\begin{eqnarray}
\overline{|\alpha_{\ell m}|^2}=\alpha(4\pi c)^2\times \frac{1}{2\ell(\ell+1)},
\end{eqnarray}
which is essential to be in accordance with the observations. It turns out that at the end of inflation $\overline{\langle\pi\rangle^2}=\lambda_c k^2 {\mathcal T}/2$, hence rescaling $\lambda_c=\tilde\lambda_c/k$ would yield $\overline{\langle\pi\rangle^2}=\tilde\lambda_c k {\mathcal T}/2$ and therefore a scale invariant spectrum. Demanding the collapse rate $\tilde\lambda_c$ now appearing in the final expression to be of dimension sec$^{-1}$, the original collapse rate parameter $\lambda_c$ should be dimensionless. 
On the other hand, choosing the collapse parameter to be the field fluctuation itself, $\hat{A}=\hat{\pi}=a\delta\hat{\phi}$, the power spectrum becomes proportional to $\lambda_c{\mathcal T}/2$, and hence to obtain a scale invariant spectrum a rescaling like $\lambda_c=\tilde\lambda_c k$ is called for. Hence, now to have $\tilde\lambda_c$ of the dimension of rate, i.e. sec$^{-1}$, one requires to demand the original collapse rate parameter $\lambda_c$ to be of dimension  sec$^{-2}$.

Knowing the $\alpha_{\ell m}$'s helps one to calculate the power spectrum of the CMB temperature anisotropies as
\begin{eqnarray}
\overline{\left(\frac{\Delta T}{T}\right)^2}=\frac{\pi}{4}\left(\frac{4\pi G\phi'_0}{3a}\right)^2\tilde\lambda_c\mathcal{TI},
\end{eqnarray}
where $\mathcal{I}$ counts the range of comoving wavenumbers which are relevant in CMB observations, which is about 10. Noting that the slow-roll parameter $\epsilon$ can be cast as
\begin{eqnarray}
\epsilon=-\frac{\dot H}{H^2}=\frac32\frac{\phi_0^{'2}}{a^2V(\phi_0)},
\end{eqnarray}
which yields the anisotropy spectrum as 
\begin{eqnarray}
\overline{\left(\frac{\Delta T}{T}\right)^2}\sim\frac{\epsilon V}{M_{\rm Pl}^4}\tilde\lambda_c\mathcal{TI},
\end{eqnarray}
and taking into account of the post reheating epochs leads to a further suppression of the power by a factor of $\epsilon^2$ which then yields the final form of the anisotropy spectrum as 
\begin{eqnarray}
\overline{\left(\frac{\Delta T}{T}\right)^2}\sim\frac{V}{\epsilon M_{\rm Pl}^4}\tilde\lambda_c\mathcal{TI}.
\label{temp-power}
\end{eqnarray}
It is to note that as the method of generating primordial inhomogeneities in this scenarios is different from the standard inflationary prescription, the methodology to calculate the observables in this method is also very different from the standard way of calculating the primordial power spectrum of the superhorizon modes and extracting the observables from that derived spectrum.
\subsection{CSL Mechanism II}

We will very briefly state the outline of this method of implementing CSL collapse dynamics within the the framework of inflation. For detailed analysis of this mechanism refer to \cite{Martin:2012pea, Das:2013qwa, Das:2014ada}. The modified Schr\"{o}dinger equation under which the wavefunctional of the wavefunctions of the mode of inflationary perturbations evolve can be written as 
\begin{eqnarray}
d\Psi_{\mathbf k}=\left[-i\hat{\mathcal H}_{\mathbf k}d\tau+\sqrt{\gamma}\left(\hat\zeta_{\mathbf k}-\langle \hat\zeta_{\mathbf k}\rangle \right)dW_\tau-\frac\gamma2\left(\hat\zeta_{\mathbf k}-\langle \hat\zeta_{\mathbf k}\rangle \right)^2d\tau\right]\Psi_{\mathbf k},
\end{eqnarray}
where the Wiener process $W_\tau$ encodes the stochastic behaviour of the CSL dynamics and $\gamma$ is the collapse parameter which has a dimension of mass$^{-2}$ or sec$^{-2}$ irrespective of whether the collapse happens in the momentum basis or in the field basis. The collapse parameter was first treated as a constant in \cite{Martin:2012pea}. The problem of treating it as a constant is that the strength of collapse would be same for all modes, be in subhorizon or superhorizon, hence even a subhorizon mode can collapse before its quantum evolution according to the standard inflationary mechanism. It is desired phenomenologically to let the modes evolve quantum mechanically while being subhorizon and then let the collapse of wavefunction happens on superhorizon scales. It is one way of implementing the `amplification mechanism' of CSL dynamics  \cite{Bassi:2012bg} of the non-relativistic system. Thus, a phenomenological model was proposed in \cite{Das:2013qwa} where the form of the collapse parameter was taken to be 
\begin{eqnarray}
\gamma=\frac{\gamma_0(k)}{(-k\tau)^\alpha},
\label{collapse-param}
\end{eqnarray}
where $\alpha$ is a positive quantity. Thus on subhorizon scales when $-k\tau$ is very large the collapse parameter would be too weak to collapse the quantum modes of primordial perturbations, and will allow them to evolve quantum mechanically while being subhorizon. As $-k\tau$ tends to vanish on superhorizon scales, the collapse parameter becomes stronger and would help the mode functions to collapse faster. The desired macro-objectification of modes happens in the parameter range $1<\alpha<2$  \cite{Das:2013qwa}. There are two issues with this phenomenological model. First of all, it yields a scale dependent power spectrum which is contrary to the observations. Secondly, the primordial modes do not `freeze' on superhorizon scales.\footnote{Both these issues are also there when $\gamma$ is treated to be a constant \cite{Martin:2012pea}.}
 To encounter the first problem it was proposed to redefine $\gamma_0(k)$ as 
\begin{eqnarray}
\gamma_0(k)=\tilde\gamma_0\left(\frac{k}{k_0}\right)^\beta,
\end{eqnarray}
with $\beta$ to be positive. This yields a nearly scale invariant spectrum if the combination $\delta=3+\alpha-\beta$ is of the order of the slow-roll parameters. To take into account of the second issue the power spectrum and hence the observables are calculated at the end of inflation. The observational imprint such a mechanism would leave would be in terms of modified scalar and tensor spectral indices and a modified consistency relation \cite{Das:2014ada}. 
\section{Fine-tuning of couplings in various theories}
\label{observables}
\subsection{The generic scenario}
To analyze why a non-minimal curvature coupling helps ease the severe fine-tuning of the self-coupling constant, we recall that the scalar power amplitude in a single field inflationary model is written as 
\begin{eqnarray}
\left.A_s\right|_{\rm single-field}=\frac{1}{24\pi^2M_{\rm Pl}^4}\frac{V}{\epsilon_V},
\end{eqnarray}
where $M_{\rm Pl}$ is the reduced Planck mass and $\epsilon_V$ is the first slow-roll parameter. Thus for a quartic potential $V(\phi)=\lambda\phi^4/4$, where $\epsilon_V=8M_{\rm Pl}^2/\phi^2$ and $N_*\simeq \phi_*^2/8M_{\rm Pl}^2,$ the amplitude would become 
\begin{eqnarray}
A_s=\frac{2\lambda}{3\pi^2}N_*^3,
\label{minimal-scalar}
\end{eqnarray}
and requiring the scalar amplitude to be of the order of $10^{-9}$, according to the observations, it calls for a severely fine-tuned quartic self-coupling $\lambda\sim 10^{-13}$. 

In the case of non-minimal coupling, the analysis takes a simple form if one makes a conformal transformation to the Einstein frame where the non-minimal coupling disappears leaving the gravity sector minimally coupled to the matter sector. In such a frame, a field redefinition makes the kinetic term of the scalar field canonical and yields an exponentially flat plateau-like potential in the large field limit $\phi\ll M_{\rm Pl}/\sqrt{\xi}$ \cite{Bezrukov:2007ep} :
\begin{eqnarray}
V\simeq\frac{\lambda M_{\rm Pl}^4}{4\xi^2}.
\label{pot-e}
\end{eqnarray}
In this large field limit with $\epsilon_V=(4M_{\rm Pl}^4)/(3\xi^2\phi^4)$ and $N_*\simeq (6\xi\phi_*^2)/(8M_{\rm Pl}^2)$, one gets the scalar spectral amplitude as
\begin{eqnarray}
A_s=\frac{\lambda}{72\pi^2\xi^2}N_*^2.
\label{non-minimal-scalar}
\end{eqnarray}
Comparing Eq.~(\ref{minimal-scalar}) and Eq.~(\ref{non-minimal-scalar}), it can be seen (which had also been observed in \cite{Fakir:1990eg} previously) that 
\begin{eqnarray}
\lambda^{(\xi>0)}=48\xi^2 N_*\lambda^{(\xi=0)},
\end{eqnarray}
hence a $\xi$ as large as $10^4$ can compensate for the smallness of the self-coupling in the minimal case to yield $\lambda$ as large as $10^{-1}$ for the non-minimal case. 
\subsection{CSL Mechanism I}

It has been stated in \cite{Rodriguez:2017rjh} that considering the inflationary scale to be near GUT scale, i.e. $V^{1/4}\sim10^{15}\,{\rm GeV}\sim 10^{-3}M_{\rm Pl}$\footnote{We will consider $M_{\rm Pl}$ as the reduced Planck mass for restating the results of \cite{Rodriguez:2017rjh}. Apart from this we will use GUT scale as $10^{15}$ GeV  consistently.}  and putting $\epsilon\sim10^{-2}$, ${\mathcal I}\sim10$ and the observed value of the amplitude of temperature anisotropy spectrum as $10^{-10}$ in Eq.~(\ref{temp-power}), one gets 
\begin{eqnarray}
\tilde\lambda_c{\mathcal T}\sim 10^{-1}.
\end{eqnarray}
To determine the conformal time $\mathcal{T}$, we know that the temperature of the Universe scales inversely to the scale factor. Hence taking the scale factor $a$ to be one today, and the temperatures today and at the end of inflation as 2.7 K=2.4$\times 10^{-13}$ GeV and $10^{15}$ GeV respectively, one can determine the scale factor at the end of inflation as 
\begin{eqnarray}
a(\tau)=\frac{T_{\rm today}}{T(\tau)}=2.4\times 10^{-28}.
\label{a-end}
\end{eqnarray}
Assuming the energy scale at the end of inflation is the GUT scale, the Hubble parameter at that time would be $H\sim 10^{11}$ GeV, following the Friedmann equation. Then as $a(\eta)=-1/H\eta$ ($\eta$ being the conformal time) during inflation, we have 
\begin{eqnarray}
|\tau|=1/aH\sim10^{17}\,{\rm GeV}^{-1}.
\end{eqnarray}
As the Hubble parameter remains (nearly) constant during inflation we have 
\begin{eqnarray}
\frac{\mathcal{T}}{\tau}=\frac{a(\tau)}{a({\mathcal{T}})}=e^{60}\sim 10^{26},
\end{eqnarray}
yielding the conformal time at the beginning of inflation $\mathcal{T}\sim10^{43}\,{\rm GeV}^{-1}\sim 10^{18}$ sec and hence the collapse rate as $\tilde\lambda_c\sim 10^{-19}$ sec$^{-1}$. 

There are quite a few points which we would like to emphasise here. First of all, the mechanism of generating primordial perturbations and obtaining the primordial observational quantities proposed by Sudarsky {\it et al.} is quite different from the standard scenario, and the striking difference between the two mechanisms is that the one proposed by Sudarsky {\it et al.} does not yield any tensor modes as the gravity in this scenario is treated classically. In the standard picture the scale of inflation is determined by the observational upper-bound on tensor-to-scalar ratio, which is at or below the GUT scale according to the present bound on $r$ \cite{Ade:2015lrj}. But it is not clear how the inflationary scale can be obtained in the proposed mechanism by Sudarsky {\it et al.} in absence of any tensor modes. Hence the assumption of inflationary scale to be at the GUT scale in the above discussion is quite ad hoc and has been borrowed from the understanding of the perturbation theory in the standard picture, which this proposed mechanism tends to overrule. Moreover, this analysis does not comment on what the quartic self-coupling of the inflaton field would be required to explain the observations in the minimal coupling scenario. Above all, while all the quantities are being calculated at the end of inflation, it would have been more appropriate to take the slow-roll parameter $\epsilon$ to be $\mathcal{O}(1)$ instead of $\mathcal{O}(10^{-2})$. 

In case of non-minimal coupling, the potential becomes exponentially flat in the Einstein frame as given in Eq.~(\ref{pot-e}). Considering $\epsilon\sim10^{-2}$, ${\mathcal I}\sim10$ and the quartic self-coupling constant $\lambda\sim10^{-1}$\footnote{In the calculation presented in \cite{Rodriguez:2017rjh} $\lambda\sim1$ has been considered.} one gets from Eq.~(\ref{temp-power}) 
\begin{eqnarray}
\frac{\mathcal{T}\tilde\lambda_c}{4\xi^2}\sim 10^{-12}.
\label{T-einstein}
\end{eqnarray}
To calculate $\mathcal{T}$ in the Einstein frame we note that the scale factors in Jordan frame and Einstein frame are related by $\tilde{a}({\tilde \eta})=\Omega(\eta) a(\eta)$, where we have calculated $a(\eta)$ at the end of inflation in Eq.~(\ref{a-end}). It is shown in \cite{Rodriguez:2017rjh} that at the end of inflation $\Omega(\tau)\sim \mathcal{O}(1)$. The Hubble parameter in the Einstein frame for this plateau-like flat potential would be 
\begin{eqnarray}
H^2=\frac{0.1}{48}\frac{M_{\rm Pl}^2}{\xi^2},
\end{eqnarray}
which gives $H\sim \frac{1}{\xi}\times10^{16}\,{\rm GeV}$. Thus the conformal time at the end of inflation would be $\tau=(\tilde{a}H)^{-1}\sim \xi\times 10^{12}\,{\rm GeV}^{-1}\sim \xi\times 10^{-13}$ sec. Then the conformal time at the beginning of inflation would be $\mathcal{T}\sim \xi e^N\times 10^{-13}$ sec, where $N$ is the number of efolds in the Einstein frame. Putting back this value in Eq.~(\ref{T-einstein}) one gets
\begin{eqnarray}
\frac{\tilde\lambda_c}{\xi}\sim 40\times e^{-N}\,{\rm sec}^{-1}.
\end{eqnarray}
Thus demanding $\xi\sim{\mathcal O}(1)$ and $N\sim 60$ one gets $\tilde\lambda_c\sim 10^{-25}$.

\subsection{CSL Mechanism II}
The scalar power amplitude, calculated at the end of inflation in the case of CSL modified dynamics, can be written as \cite{Das:2014ada}
\begin{eqnarray}
\left.A_s\right|_{\rm CSL}=\left(\frac{k_0^2}{\tilde{\gamma}_0}\right)\left(\frac{k_*}{k_0}\right)^\delta e^{-(1+\alpha)N_*}\times \left.A_s\right|_{\rm single-field}.
\end{eqnarray}
We can see from this equation that the CSL dynamics modified scalar amplitude contains three parameters of the collapse theory, $\tilde\gamma_0$, $\alpha$ and $\delta$, where $\delta$  is a combination of two other parameters of the theory $\alpha$  and $\beta$. As $\delta$ also appears in the scalar spectral index expression, it can be at best of the order of the slow-roll parameters $(\mathcal{O}(10^{-2}))$, if not a null number. The parameter $\alpha$ is positive by construction and should lie within $1<\alpha< 2$ to yield the observed macro-objectifications of the primordial modes. On the other hand, there exists no such bound of the strength parameter $\tilde\gamma_0$ of the collapse model, which awaits a formulation of  a field theoretic version of such a quantum collapse dynamics. Thus, the order of magnitude contribution to the scalar power of the first term on the rhs of the above equation is not known. The second term contains two comoving length scales, $k_0=H_0=(4400\,{\rm Mpc})^{-1}$  (with $a_0=1$) and $k_*=0.002\,{\rm Mpc}^{-1}$, and with $\delta\sim 10^{-2}$ the term contributes at the order of unity to the scalar power. The consecutive term, on the other hand, can exponentially suppress the scalar power, e.g. for $\alpha\sim1.1$ and $N_*\sim 60$, this suppresses the power at $\mathcal{O}(10^{-55})$. The larger the value of $\alpha$ the more is the suppression of power. 

Now, if the collapse dynamics is applied to the minimally coupled model, then to obey the present observations of CMB fluctuations the quartic self-coupling should be 
\begin{eqnarray}
\lambda\sim10^{-9}\times \frac{3\pi^2}{2N_*^3}\times e^{(1+\alpha)N_*}\left(\frac{\tilde{\gamma}_0}{k_0^2}\right)\sim 10^{41} \left(\frac{\tilde{\gamma}_0}{k_0^2}\right).
\end{eqnarray}
Hence for $\lambda\sim 0.1$, one needs to fine-tune the collapse strength parameter to the order $10^{-41}$ in units of $k_0^2$. The smaller the value of the self quartic coupling $\lambda$ the more the fine-tuning required for the strength parameter $\tilde\gamma_0$. But as this kind of collapse dynamics renders the same tensor-to-scalar ratio, the scenario would still yield too large $r$ to be in accordance with observations. Hence such a model is certainly at odds with the current cosmological observations. 

If we now turn to a scenario where collapse dynamics is applied to the non-minimally coupled scenario, then the quartic self-coupling should be
\begin{eqnarray}
\lambda\sim10^{-9}\times \frac{72\pi^2}{N_*^2}\times e^{(1+\alpha)N_*}\xi^2\left(\frac{\tilde{\gamma}_0}{k_0^2}\right)\sim 10^{45} \xi^2\left(\frac{\tilde{\gamma}_0}{k_0^2}\right).
\end{eqnarray}
We must recall here that the Higgs inflation scenario by construction is devised in the large field limit $\phi\ll M_{\rm Pl}/\sqrt{\xi}$ which helps yield a plateau-like potential. Hence a very small non-minimal coupling would call for inflaton field values to be extremely super-Planckian, which is undesired. Thus one can, at best, expect $\xi\sim\mathcal{O}(1)$, or $\xi=-1/6$, if lower, which is a well-motivated value dubbed the conformal coupling. In such a case, $\tilde\gamma_0$ to be fine tuned to $10^{-46}$ (for $\xi\sim\mathcal{O}(1)$) or $10^{-44}$ (for $\xi\sim-1/6$) in units of $k_0^2$. Such fine tuning of the parameter $\tilde\gamma_0$ should not raise concern, as this negligible collapse parameter ensures quantum evolution of the sub-horizon modes. Only on super-horizon scales when the collapse parameter $\gamma$ tends to grow stronger, the quantum modes tend to behave classically (see Eq.~(\ref{collapse-param})). 

\section{Discussion and Conclusion}
\label{conclusion}

We have addressed the issue of `unitarity problem' or the `naturalness problem' of Higgs inflationary scenario within the framework of quantum collapse dynamics, which has been recently employed in the inflationary dynamics in order to explain the longstanding `interpretational issue' of quantum to classical transition of primordial perturbations. A similar work has been recently done by Rodriguez and Sudarsky in \cite{Rodriguez:2017rjh}. As there are two different approaches through which the CSL collapse dynamics is integrated into inflationary mechanism, one formulated by Sudarsky and collaborators  \cite{Canate:2012ua, Landau:2011aa} and the other one devised by Martin and collaborators \cite{Martin:2012pea} and later developed in \cite{Das:2013qwa, Das:2014ada, Banerjee:2016klg}, we have analysed both the approaches in this article to see whether the collapse mechanism helps alleviate the `unitarity problem' of Higgs inflation scenario by taming the required large value of the non-minimal curvature coupling. 

We noticed that both these approaches can indeed help bring down the large value of $\xi$ which is the main cause behind the `unitarity problem' of Higgs inflation. It might appear that as the collapse dynamics brings in new parameters in the theory, like the collapse parameter, the fine-tuning of $\xi$ can now be relaxed by transforming that to the newly introduced parameter. But we noticed, that the constraint on $\xi$ is being relaxed because in both these approaches the observables are being calculated at the end of inflation, which suppresses the power spectrum by inverse of exponential of the e-foldings of inflation. The reason of calculating the observables at the end is different in these two different approaches. The first approach proposed by Sudarsky and collaborators calculates them at the end of inflation due to the assumption that the primordial inhomogeneities and anisotropies are imprinted on the matter sector at the end of inflation. In the second approach the primordial power spectrum is calculated at the end of inflation because due to the evolution under the collapse dynamics the mode functions on superhorizon scales do not `freeze', and hence it cannot be calculated at the time of horizon-crossing of these modes. 

It is important to note that in \cite{Canate:2012ua, Rodriguez:2017rjh} Sudarsky and collaborators have tend to compare the value of the collapse parameter $\tilde\lambda_c$ with its non-relativistic counterpart where it is expected that for quantum mechanical systems the collapse parameter would be within the range of $10^{-10}$ to $10^{-20}$. In such a case, to obtain $\xi\sim{\mathcal{O}}(1)$ in this mechanism one requires $\tilde\lambda_c\sim 10^{-25}$, which does not lie in the desired range. On the other hand, if one insists to have the value of collapse parameter within the desired range of its non-relativistic counterpart, then one has to compromise with the duration of inflation which can be within 30 to 50 efolds \cite{Rodriguez:2017rjh}. But in the second approach, the collapse parameter for the relativistic fields is not related to the one for the non-relativistic systems, hence the bounds obtained from the laboratory systems are not applicable to this relativistic systems \cite{Martin:2012pea}. We showed that a fine-tuned collapse parameter $\tilde\gamma_0\sim10^{-44}$ can accommodate both $\gamma\sim0.1$ and $\xi\sim\mathcal{O}(1)$ which help evade the concern of the `unitarity problem' in Higgs inflation scenario.

The distinguishable feature of which of these approaches are correct depends on more precise observations of spectral indices and the tensor-to-scalar ratio. For the first approach the tensor-to-scalar ratio and the tensor spectral index both vanish, though this approach does not comment on the form of the scalar spectral index. In the second approach the tensor-to-scalar ratio would be of the order of $10^{-3}$, just like in the standard Higgs inflation scenario, while both the spectral indices, scalar and tensor, and consistency relation of single-field inflationary model would be modified by the collapse parameter $\delta$ \cite{Das:2014ada, Banerjee:2016klg}. 

\begin{acknowledgments}
 Work of SD is supported by the Department of Science and
 Technology of Government of India, under the Grant Agreement number IFA13-PH-77 (INSPIRE Faculty
 Award).  The author would like to thank K Sravan Kumar for useful discussions. 
 \end{acknowledgments}

\bibliographystyle{JHEP}
\providecommand{\href}[2]{#2}\begingroup\raggedright\endgroup

\end{document}